\documentclass[12pt]{iopart}
\usepackage{graphicx}
\usepackage{subfigure}

\begin{document}

\title{Simulation studies of permeation through two-dimensional ideal polymer
networks}
\author{Yong Wu, B Schmittmann and R K P Zia}

\begin{abstract}
We study the diffusion process through an ideal polymer network, using
numerical methods. Polymers are modeled by random walks on the bonds of a
two-dimensional square lattice. Molecules occupy the lattice cells and may
jump to the nearest-neighbor cells, with probability determined by the
occupation of the bond separating the two cells. Subjected to a
concentration gradient across the system, a constant average current flows
in the steady state. Its behavior appears to be a non-trivial function of
polymer length, mass density and temperature, for which we offer qualitative
explanations.
\end{abstract}

\maketitle

\address{Department of Physics, Virginia Polytechnic Institute and State University, Blacksburg, VA
24061-0435} \ead{wuyong@vt.edu}

\section{Introduction}

The permeation process through a polymer network attracted much recent
attention among both experimentalists and theorists.\cite
{KeFr93,LaMaScZi03,ScGoZi05} Experiments can be easily conducted, by
applying a constant gas pressure across a thin polymer film so that the gas
molecules are forced to permeate through it. The current at the steady state
are measured, whose value is dependent on the temperature, the size of gas
molecules, and the fraction and distribution of free volume, etc.\cite
{ThKrFa99} In order to understand this phenomenon theoretically, a
lattice-gas model has been constructed, in which the polymers are modeled by
random walks on the bonds of a lattice, and the gas molecules are placed in
the lattice cells.\cite{ScGoZi05} Gas molecules can diffuse to the neighbor
cells by jumping across the bonds. If a bond is occupied by one or more
polymer segments, the jump only happens when the molecule can overcome an
energy barrier. In a simulation, a constant gradient of gas molecule
concentration is maintained on two opposite sides of the lattice, producing
numerically measurable gas current. The effective diffusivity is simply
proportional to the current.

The lattice-gas model for gas permeation is mathematically equivalent to the
well-known random resistor network model. In previous studies of random
resistor networks, however, it is usually assumed that the probability of a
bond being occupied by a resistor is independent of other bonds.%
\cite{DeVa82,HaLu87,Luck91} Our work can thus also be deemed as a
study of the random resistor network with correlations, since a
polymer induces non-trivial correlations between its segments (the
occupied bonds in this mapping).

In this paper we are particularly interested in the relation between the
current at the steady state (or the effective diffusivity) and the polymer
length. The polymer length can be considered as a parameter that controls
the strength of the correlations. When the polymer length grows, the spatial
distribution of energy barriers has more fluctuations. Therefore both the
``free volume'' where molecules can move freely, and the height of energy
barriers increase. These two factors compete against each other, which
results in a complicated behavior of the current as a function of the
polymer length. In this paper we numerically measure the current for
different polymer lengths and give qualitative explanations to understand
the behavior.

The layout of this paper is as follows. We describe our model in details in
section \ref{sec:model}. The simulation results are shown in section \ref
{sec:simulation}, and qualitative explanations to the results are also
given. We end with a summary and outlook.

\section{Model}

\label{sec:model} Computer simulations of polymers can be quite
detailed and sophisticated. In this paper we only use a very
coarse-grained model as the first step towards the theoretical
understanding of the problem of gas permeation through a polymer
network. We model the polymers by random walks on a two-dimensional
square lattice. The random walks can either be completely
non-interacting, generating bonds that are multiply occupied by
polymer segments, or they may follow the rules that each bond can
only be occupied by a single polymer segment. The gas molecules are
located inside the lattice cells. For each sweep of the simulation,
each molecule attempts a jump to one of the neighbor cells. The
attempt is accepted with unit probability if the bond to be crossed
is unoccupied by polymer segments. If the bond is occupied, then
there is an energy barrier for the molecule to overcome, and the
jumping probability is less than unity.

The overall density of of polymers is controlled by the mass density $\rho $%
,
\begin{equation}
\rho =\frac{M\ell }N,  \label{eq:mden}
\end{equation}
where $M$ is the total number of the polymers thrown into the system, $\ell $
is the polymer length (i.e., number of segments in each polymer), and $N$ is
the number of all lattice bonds. Another parameter that can be used is the
occupation probability $p$, the probability of a lattice bond being occupied
by one or more polymer segments,
\begin{equation}
p=\frac{N_o}N,  \label{eq:op}
\end{equation}
where $N_o$ is the number of occupied bonds. For polymers with
exclusion, only single occupation is allowed and $p$ is simply equal
to $\rho $. If the polymers are non-interacting, multiple occupancy
of bonds means that $p$ is not a fixed quantity, differing from one
realization of the polymer configuration to another. Its average can
be related to $\rho =M\ell /N$ through $p=1-(A_\ell )^M$, where
$A_\ell $ is the average density of unoccupied bonds for a
\emph{single }random walk of length $\ell $. In the large $N$ limit,
$A_\ell \rightarrow 1-b_\ell /N$ , where $b_\ell $ is the average
number all occupied bonds in an infinite lattice \cite{AnHiZi02}.
Thus, $p\rightarrow 1-\exp \left( -\rho b_\ell /\ell \right) $ as $%
N,M\rightarrow \infty $ with fixed $\rho ,\ell $.

Molecules are driven to flow by a concentration gradient, generated
by the following boundary conditions. On a square lattice, we impose
periodic boundary conditions across the ``vertical'' direction. All
molecules enter the system on the left boundary (first column), and
whenever a molecule leaves a cell of this column, the cell is filled
with a new molecule. The right boundary (last column) is the exit,
where molecules are removed from the system whenever they reach it.
The quantity we measure is the overall current, defined as the
number of molecules exiting from the system per unit time.

Let $c(i)$ denote the average concentration of molecules at cell $i$ in the
steady state, i.e., $\left\langle n\left( i\right) \right\rangle $. where $%
n\left( i\right) =1,0$ if the cell is occupied or empty. Due to the
conservation of gas molecules in this state, we have, for each $i$,
\begin{equation}
\sum_j\left\langle t(j,i)n(j)\left[ 1-n\left( i\right) \right]
-t(i,j)n(i)\left[ 1-n\left( j\right) \right] \right\rangle =0,  \label{cc}
\end{equation}
where the summation runs over all cells nearest neighbor to $i$, and $t(i,j)$
is the probability of accepting a jump from cell $i$ to cell $j$. In our
study this probability is determined by the temperature $T$ and $m(i,j)$,
the number of polymer segments on the $i$-$j$ bond:
\begin{equation}
t(i,j)=q^{m(i,j)};\quad q\equiv \exp (-B/T)  \label{eq:passp}
\end{equation}
where $B$ is some constant. Thus,
\[
t(i,j)=t(j,i)
\]
leading to the cancelation of the terms in Eq.\ \ref{cc} that are
quadratic in $n$. The result is a much simpler equation:
\begin{equation}
\sum_jt(i,j)\left[ c(j)-c(i)\right] =0.  \label{eq:ssd}
\end{equation}

It is easy to see that Eq.\ \ref{eq:ssd} has the same form as the
Kirchoff's voltage law. The concentration $c(i)$ plays the role of
the electric potential at node $i$ in the dual lattice, and $t(i,j)$
is analogous to the resistance between node $i$ and node $j$.
Therefore the system at the steady state is equivalent to the
resistor network. Unlike the random resistor network that has been
much studied for years, however, the
strengths of our resistors in our system are highly correlated through $%
m(i,j)$.

\section{Simulation methods and results}

\label{sec:simulation} We consider the current as a function of the
polymer length at some typical mass densities and temperatures.
Simulations are done in a $64^2$ square lattice at $q=0.01$ and
$0.1$, representing the low temperature and high temperature regime,
respectively. To find the steady state we disregard the first $10^5$
Monte Carlo steps(MCS) to allow the system to approach the steady
state. Then, we average the cell occupation variables over $10^5$
MCS and use the resulting $c(i)$ as the initial condition to
iteratively solve Eq.\ \ref{eq:ssd}. The overall current $J$
measured as the number of exiting molecules in $10^3$ MCS. For each
temperature and each mass density (occupation probability), we
measure the current for $1000$ realizations of polymer configuration
and take the average as the result.

\begin{figure}[tbp]
\begin{center}
\subfigure[]{\label{fig:poll}\includegraphics[width=2.8in]{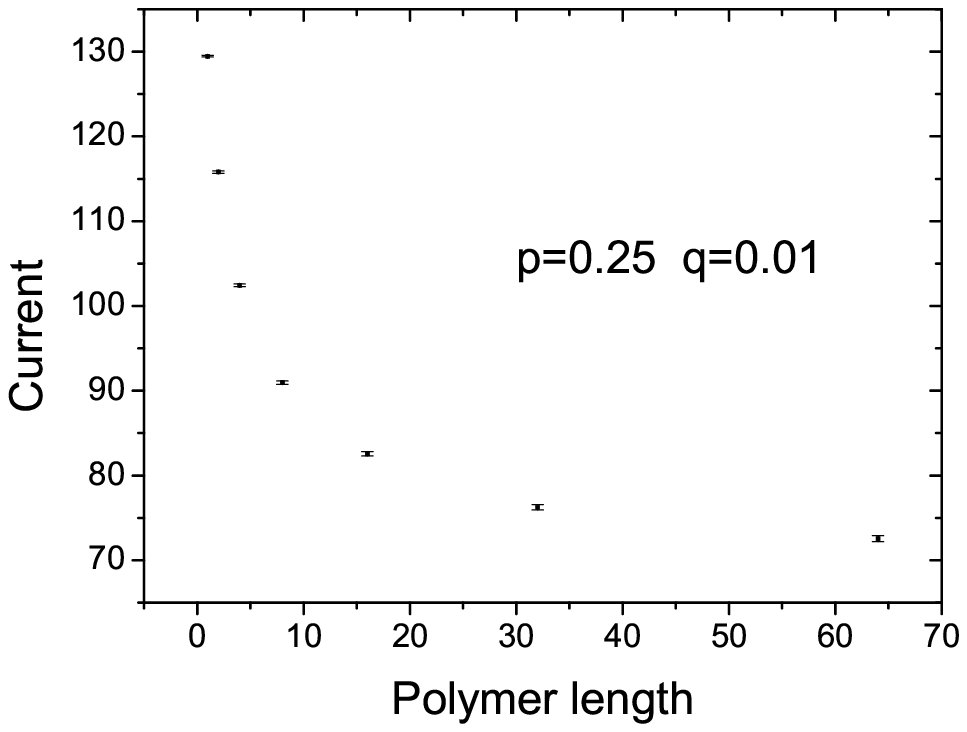}} %
\subfigure[]{\label{fig:polh}\includegraphics[width=2.8in]{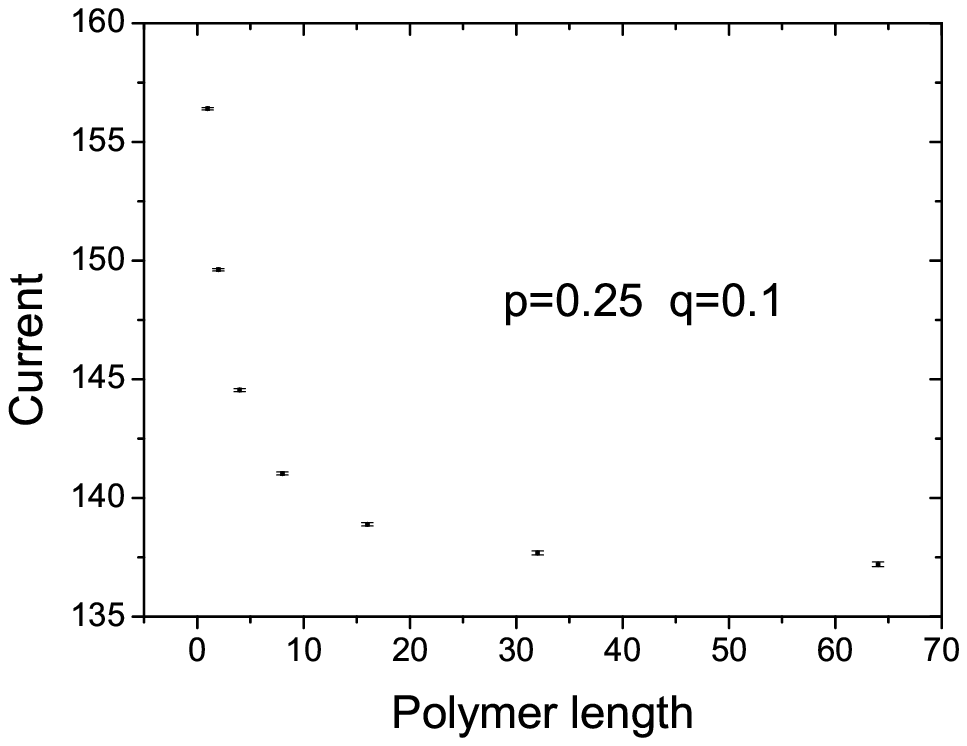}} %
\subfigure[]{\label{fig:pohl}\includegraphics[width=2.8in]{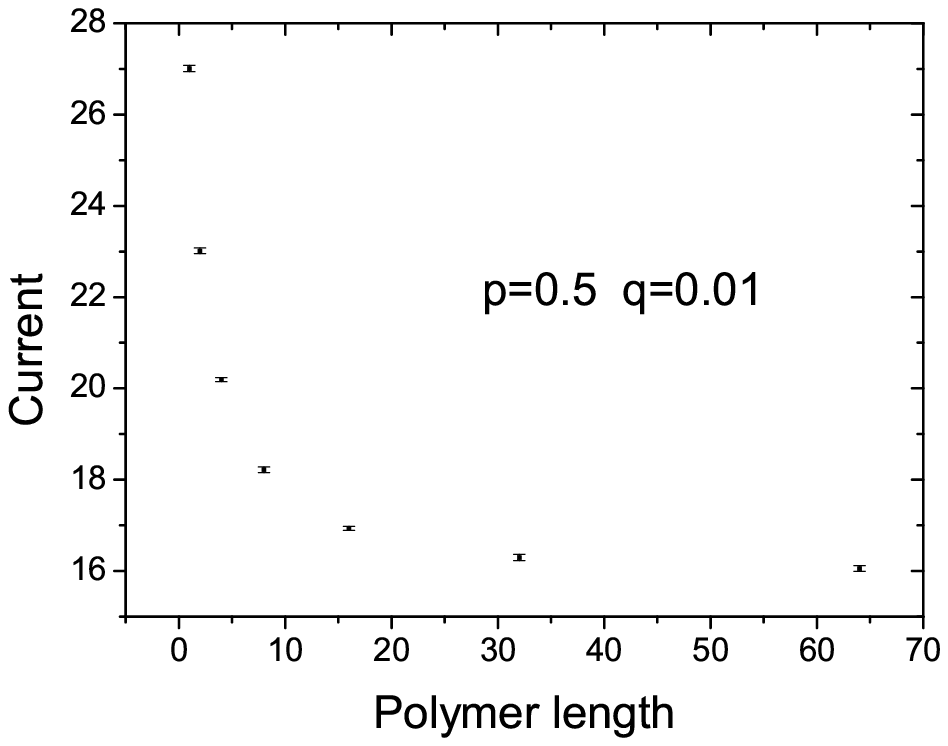}} %
\subfigure[]{\label{fig:pohh}\includegraphics[width=2.8in]{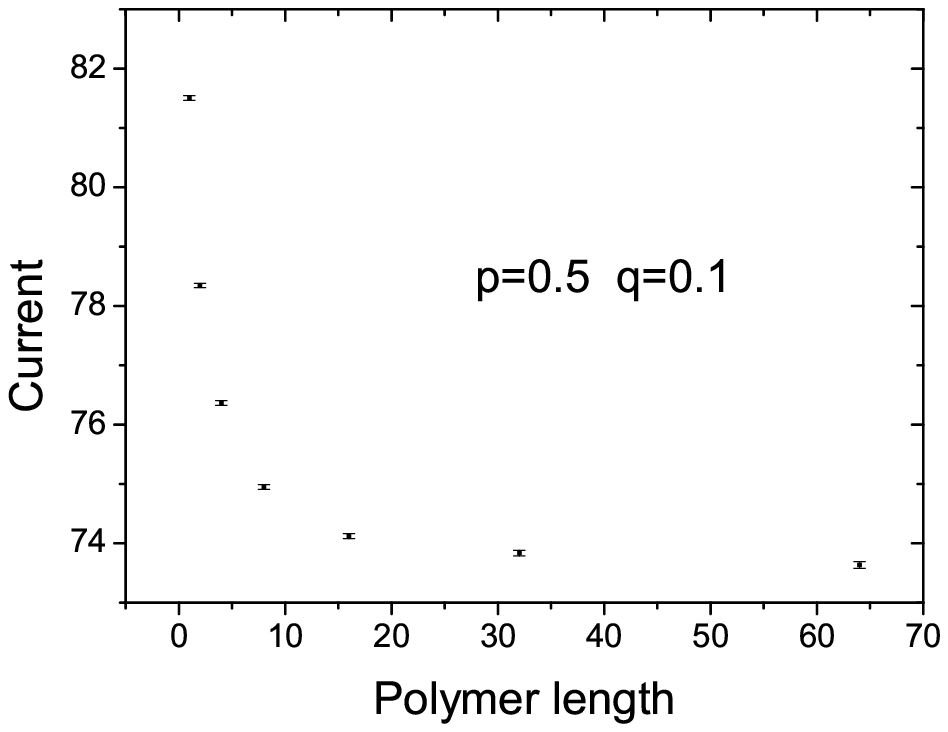}}
\end{center}
\caption{Steady state current through the polymer network as a function of
the polymer length when only single occupation of bonds is allowed. The
probability to jump across a single polymer segment $t$ and the occupation
probability $p$ are shown in the inset of each diagram.}
\label{fig:po}
\end{figure}

For simplicity we first allow only single occupancy of polymer segment on
lattice bonds. In Fig.\ \ref{fig:po} we show the current at two different
occupation probabilities, $p=0.25$ representing the low occupation, and $%
p=0.5$ representing high occupation that is close to percolation.
The relation between the current and the polymer length is quite
simple. We always observe decreasing current with growing polymer
length, which seems to approach a constant at the large $\ell$
limit. A somewhat quantitative discussion of the relation between
the current and the temperature will be given
elsewhere.\cite{WuEMT07}

\begin{figure}[tbp]
\begin{center}
\subfigure[]{\label{fig:bloblp}\includegraphics[width=2.8in]{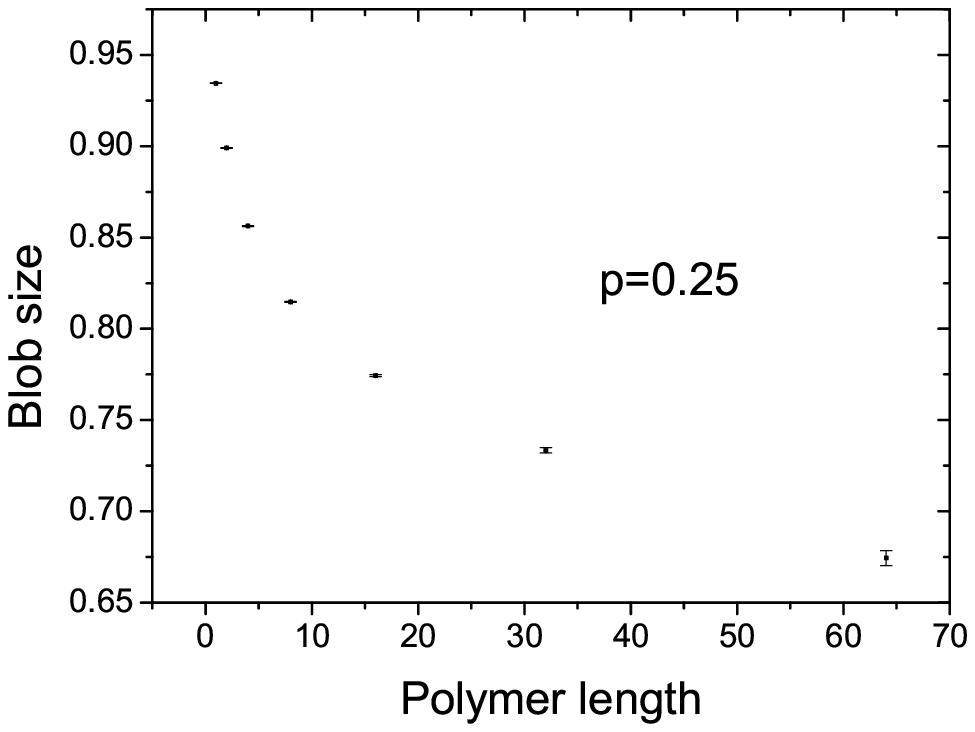}} %
\subfigure[]{\label{fig:blobhp}\includegraphics[width=2.8in]{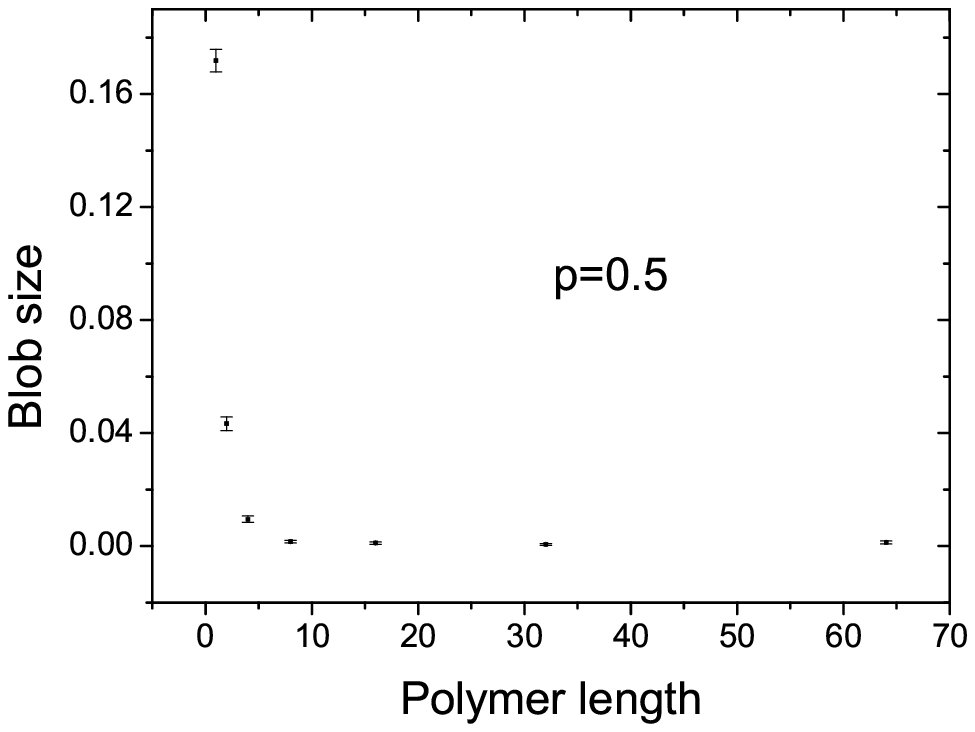}}
\end{center}
\caption{The blob size as a function of the polymer length- for the case
with \emph{single} occupation of bonds.}
\label{fig:bpo}
\end{figure}

One qualitative explanation to the observation is to consider the
blob size. At zero temperature, only a part of the cells can carry
current, which consist of the ``blob'', or ``backbone'' of the
system.\cite {ShKlReSt79,HeHoSt84} When the occupation probability
is fixed, the average blob size is a decreasing function of the
polymer length, as seen in Fig.\ \ref{fig:bpo}. At low density, it
is easier for longer polymers to surround an area and exclude it
from the backbone. When the density is high, the longer polymer
networks reach percolation first, since the percolation threshold
$p_c$ is a decreasing function of the polymer length $\ell $.
Therefore systems with longer polymer length have more fraction of
closed configurations\ (i.e., ones with zero blob size), and the
average blob size of the system is lower. The blob or the backbone
is defined at zero temperature, and plays less important role when
the temperature is high.

\begin{figure}[tbp]
\begin{center}
\subfigure[]{\label{fig:moll}\includegraphics[width=2.8in]{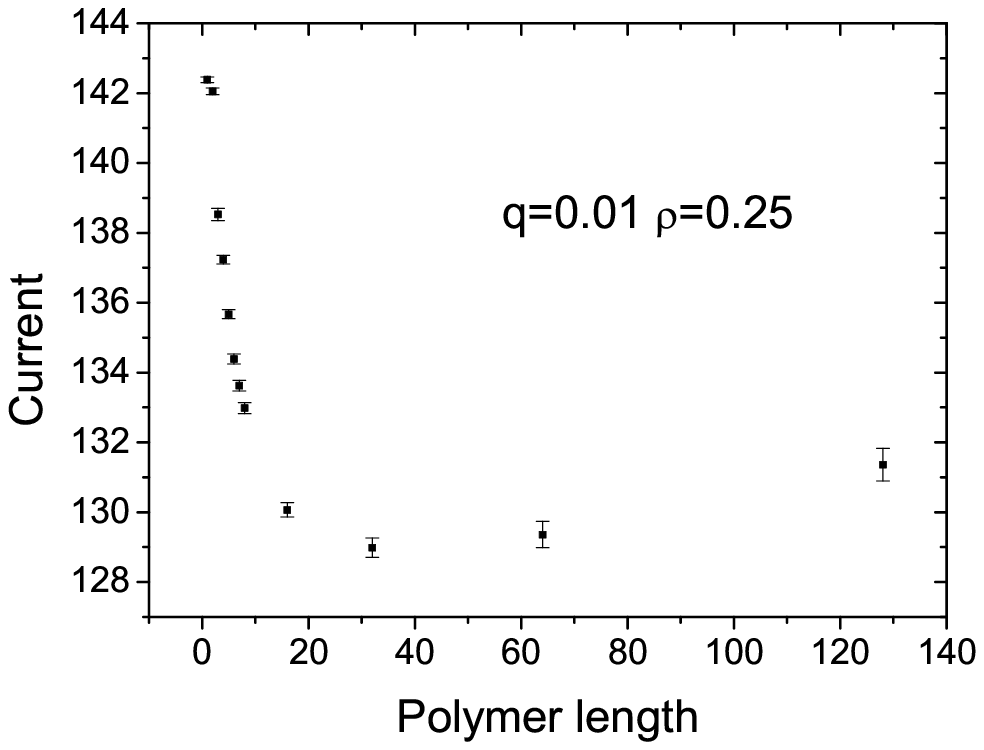}} %
\subfigure[]{\label{fig:molh}\includegraphics[width=2.8in]{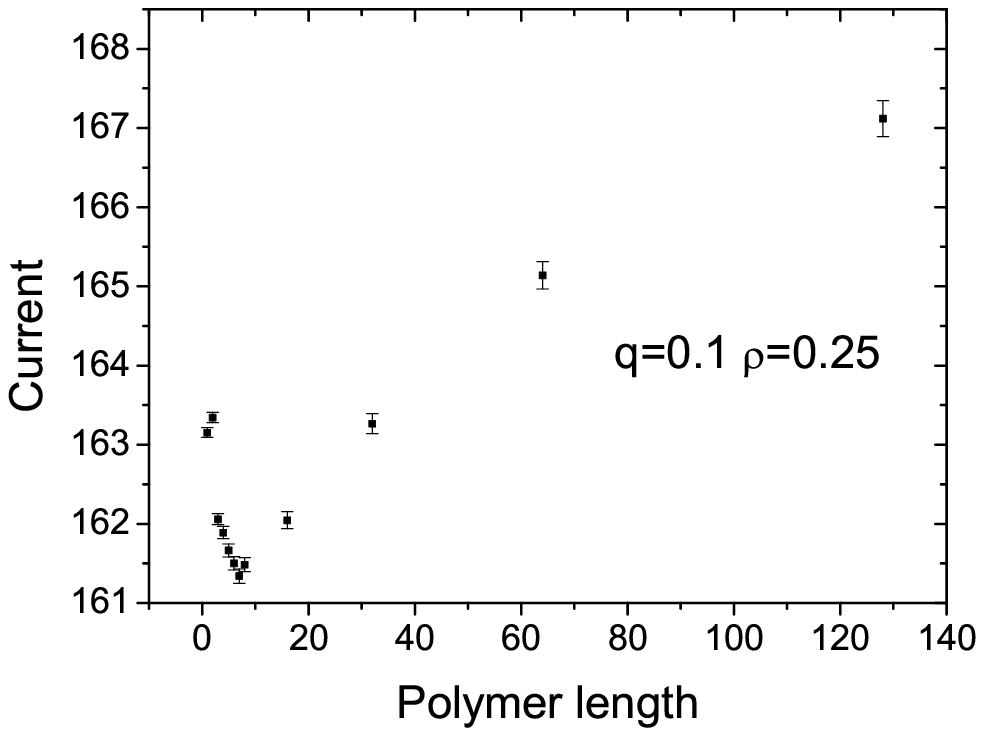}} %
\subfigure[]{\label{fig:mohl}\includegraphics[width=2.8in]{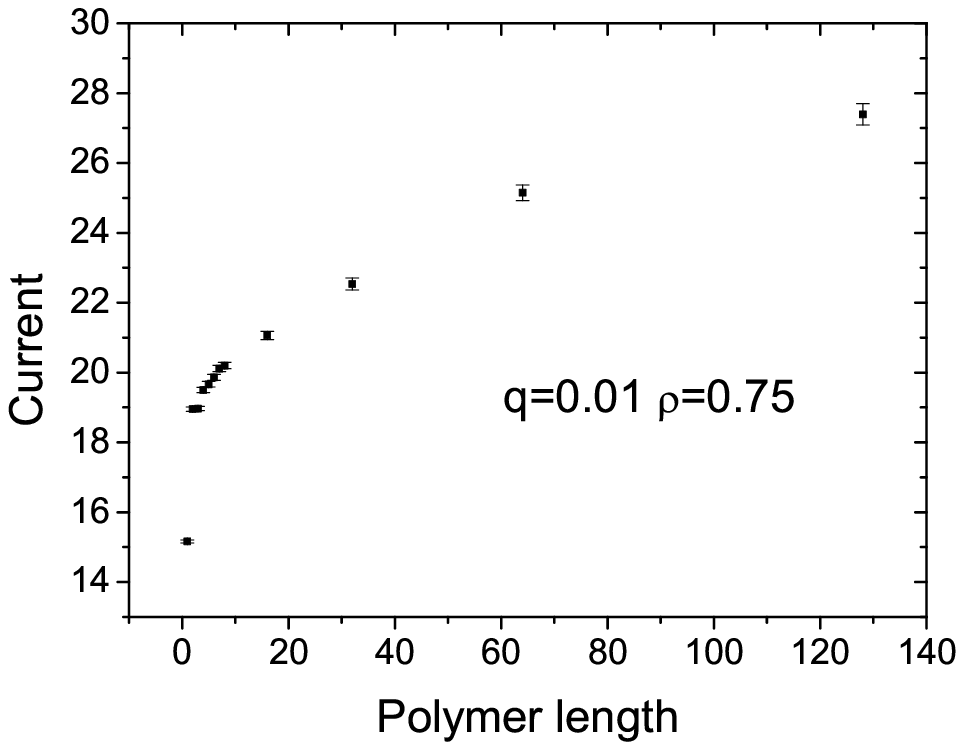}} %
\subfigure[]{\label{fig:mohh}\includegraphics[width=2.8in]{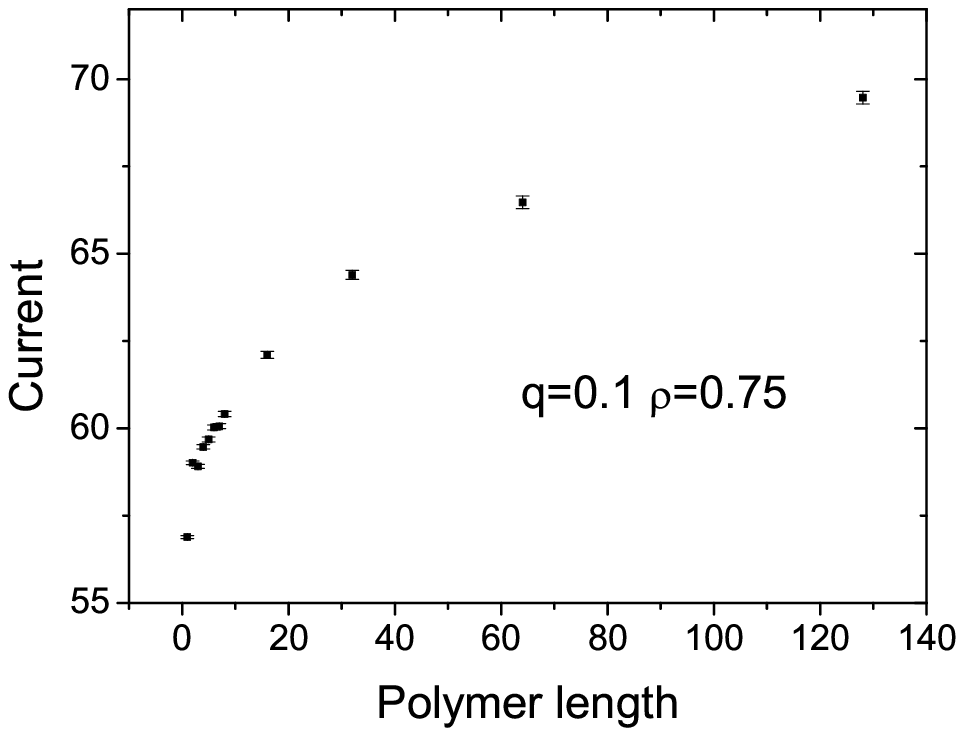}}
\end{center}
\caption{Steady state current through the polymer network as a function of
the polymer length when multiple occupation of bonds is allowed. The
probability to jump across a single polymer segment $t$ and the mass density
$\rho$ are shown in the inset of each diagram.}
\label{fig:mo}
\end{figure}

Next, we allow multiple occupation and fix the mass densities at $\rho =0.25$
and $0.75$. The former is considered to be low and far away from
percolation, while the latter is high and close to the percolation
threshold. The simulation results are shown in Fig.\ \ref{fig:mo}.

The relation between the current and the polymer length becomes rather
complicated here. In Fig.\ \ref{fig:moll} we can see that at low temperature
and low mass density, the current in general decreases when the polymer
length grows, but it also undergoes a slight increase when the polymer
length is very long. At the higher temperatures this increasing trend for
long polymers becomes more pronounced, as seen in Fig.\ \ref{fig:molh}. When
the mass density is high, however, the current is always a increasing
function of the polymer length, as shown in Fig.\ \ref{fig:mohl} and Fig.\
\ref{fig:mohh}.

\begin{figure}[tbp]
\begin{center}
\subfigure[]{\label{fig:blobl}\includegraphics[width=2.8in]{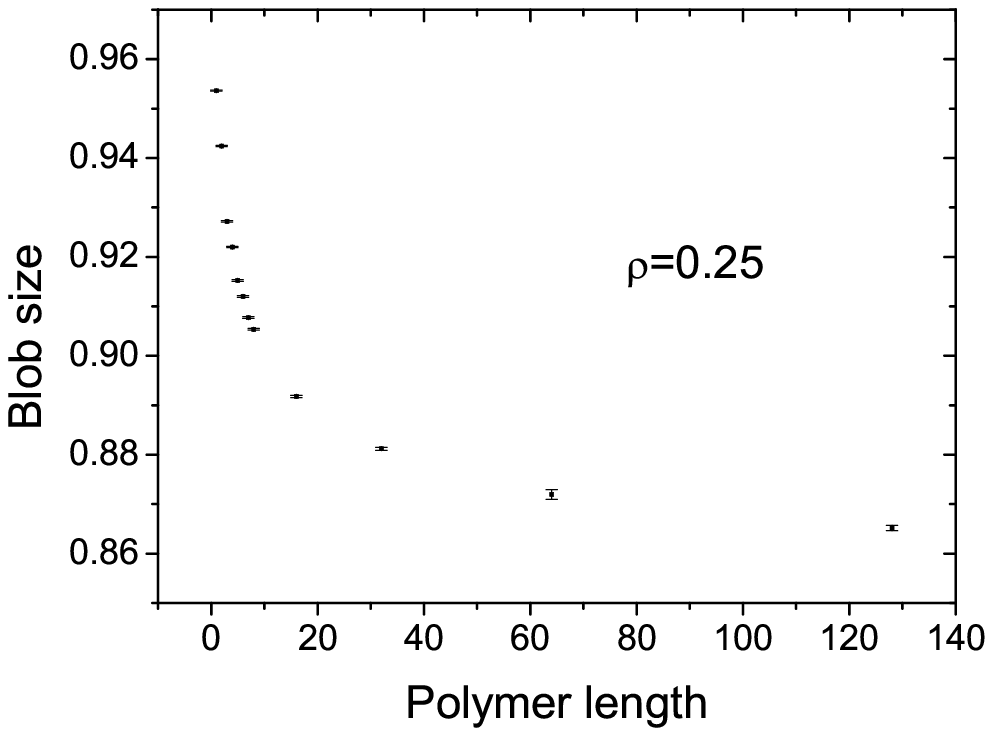}}
\subfigure[]{\label{fig:blobh}\includegraphics[width=2.8in]{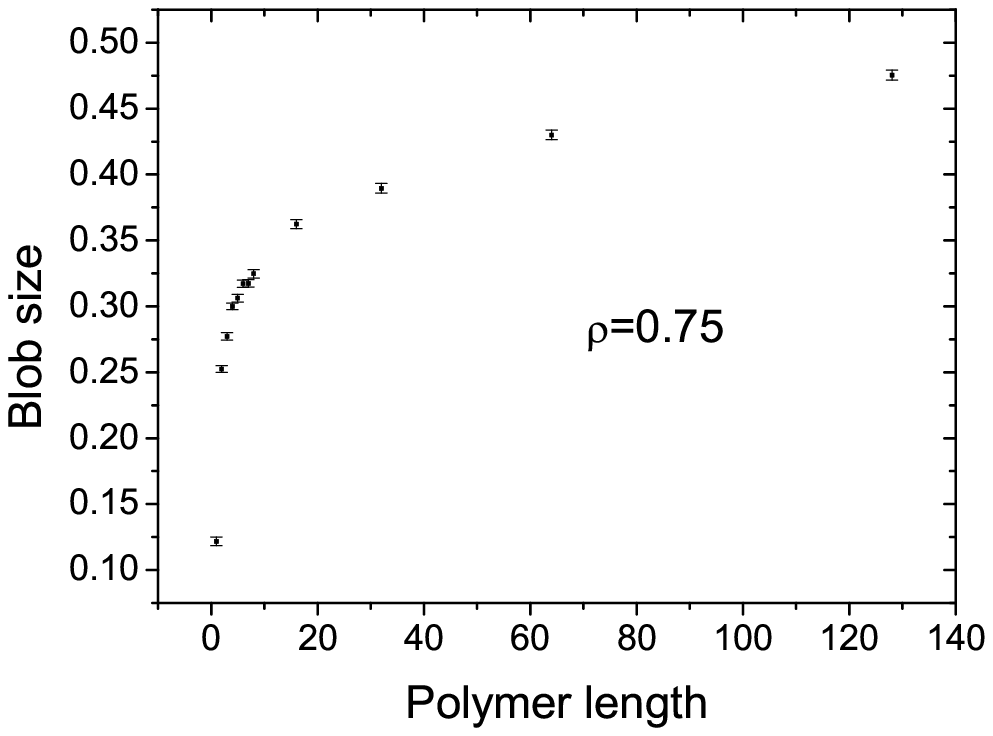}}
\end{center}
\caption{The blob size as a function of the polymer length- for the case
with \emph{multiple} occupation of bonds.}
\label{fig:bmo}
\end{figure}

For system with multiple occupation but fixed mass density, there
are at least three factors affecting the current. First, when
multiple occupation is allowed, longer polymers have lower
occupation probability, which tends to increase the current.
Secondly, fixed mass density requires that system with lower
occupation probability must have more multiple occupied bonds, and
multiple occupied bonds locally allow smaller current to get through
than singly occupied ones. Thirdly, the dependence of the average
blob size on the polymer length. The complicated behaviors of the
current result from the competition among these three factors. At
density $\rho=0.25$, the average blob size is a decreasing function
of the polymer length, as shown in Fig.\ \ref{fig:blobl}, which
dominates the behavior of the current at the lower temperature
$q=0.01$ (Fig.\ \ref{fig:moll}). However, the decreasing occupation
probability has shown its influence when $\ell\geq64$, where the
current starts to increase slightly. The impact of this factor is
seen more clearly at the higher temperature $q=0.1$ (Fig.\ \ref
{fig:molh}), since its competitor, the decreasing blob size, plays a
less significant role at higher temperatures. Though decreasing at
short polymer lengths, the current undergoes an apparent and sudden
increase when $\ell\geq8$, forming a kink at $\ell=7$. We know
little about the position of the kink except that it decreases as a
function of temperature. At the higher density $\rho=0.75$, the
behavior of the average blob size changes significantly, as shown in
Fig.\ \ref {fig:blobh}. At this density, systems with longer polymer
length has larger blob size, which, along with the decreasing
occupation probability, make the current a increasing function of
the polymer length even at low temperature (Fig.\ \ref{fig:mohl}).

\section{Summary and Outlook}

\label{sec:sum} In this paper we have applied a simple model to the
study of the permeation though a polymer network, in particular, the
relation between the steady state current and the polymer length. We
have used the Monte Carlo simulations on a two-dimensional square lattice with size up to $%
64^2$ to show that, for systems without multiple occupations, the
steady state current decreases for longer polymer lengths, which can
be related to the the average blob size of the system. When multiple
occupation is present, the current as a function of the polymer
length has a complicated behavior, which can be qualitatively
explained by the interplay between the fraction of occupied bonds
and the blob size. For properties of random media, a powerful tool
for theoretical understanding is the effective medium
theory\cite{Choy}. In particular, it generates quite accurate
results for random resistor networks, when the system is far away
from percolation \cite{Luck91}. Application of this theory to our
problem encountered some success and will be reported elsewhere
\cite{WuEMT07}.

Obvious directions for further study are (i) larger system sizes,
which should provide some insight on thermodynamic limits and
scaling properties; (ii) more detailed scan of the space of
parameters $\rho $ and $T$ . We are also aware that results from
this model cannot be easily compared with data from physical
experiments. In particular, the model needs to be generalized to
three dimensions. One possibility is to consider a cubic lattice,
with the gas molecules still occupying the cells. To model the
barriers formed by the polymers, we can let a chain be a sequence of
random steps across the \emph{diagonals} of the \emph{faces} of the
cubes. This way, a face can be considered as empty or occupied (by
one more segments), leading to free diffusion of the particle
between the adjoining cells, or an appropriately suppressed
transition rate. Beyond a simple study of the currents due to a
randomly created polymer network, we will need to introduce a model
for the dynamics of the polymers. Otherwise, we cannot study the
interesting effects of a polymer film prepared in a variety of
manners or allowed to age over time. The main obstacle for such a
program is obvious: computational power. In this sense, our study
should be considered as a first step of a long program towards a
full understanding of the complex problem of gas permeation through
polymer networks.

\section{Acknowledgment}

This research is supported in part by the US National Science Foundation
through grant DMR-0414122.

\end{document}